\documentclass[manuscript]{acmart}

\usepackage{graphicx}
\usepackage{booktabs}
\usepackage{tabularx}
\usepackage{longtable}
\usepackage{placeins}
\usepackage{float}
\usepackage{needspace}
\usepackage{multirow}
\microtypesetup{expansion=false}
\AtBeginDocument{%
  }

\setcopyright{none}
\renewcommand\footnotetextcopyrightpermission[1]{}

\begin{document}

\title{Decomposing the Associative Chain: Schema-Based Mediation Supports Expression and Understanding in Human-AI Music Co-creation}

\author{Songruoyao Wu}
\email{wsry@zju.edu.cn}
\affiliation{%
  \institution{Zhejiang University}
  \city{Hangzhou}
  \country{China}
}

\author{Wanying Wu}
\email{3240100265@zju.edu.cn}
\affiliation{%
  \institution{Zhejiang University}
  \city{Hangzhou}
  \country{China}
}

\author{Shuyu Li}
\email{lsyxary@zju.edu.cn}
\affiliation{%
  \institution{Zhejiang University}
  \city{Hangzhou}
  \country{China}
}

\author{Chongjun Zhong}
\email{zhong\_chongjun@zju.edu.cn}
\affiliation{%
  \institution{Zhejiang University}
  \city{Hangzhou}
  \country{China}
}

\author{Ziyi Huang}
\email{ziyihuang1016@zju.edu.cn}
\affiliation{%
  \institution{Zhejiang University}
  \city{Hangzhou}
  \country{China}
}

\author{Kejun Zhang}
\email{zhangkejun@zju.edu.cn}
\affiliation{%
  \institution{Zhejiang University}
  \city{Hangzhou}
  \country{China}
}

\renewcommand{\shortauthors}{Wu et al.}

\begin{abstract}
Users cannot reliably translate their creative intents into AI-generated music. We contend that the root of this difficulty lies in associative divergence: associations between narratives and musical features are overly remote, and users lack a visible mediation structure to negotiate them. To address this, we introduce five core schema groups as cognitive mediation and propose a bidirectional schema-based framework. Its forward pathway decomposes narrative-to-music associations into staged mappings, while its reverse pathway provides structured interpretations from music back to narrative. Findings from two user studies($N = 188$; $N = 60$) show that the framework led to clearer creative expression and greater perceived control, improved the comprehensibility of generated music, especially for unconventional pieces. Furthermore, it supported cultural and affective thinking while helping users locate associative divergence. Our work points toward better creative intent alignment in human-AI music co-creation.
\end{abstract}


\begin{teaserfigure}
  \centering
  \includegraphics[width=0.9\textwidth]{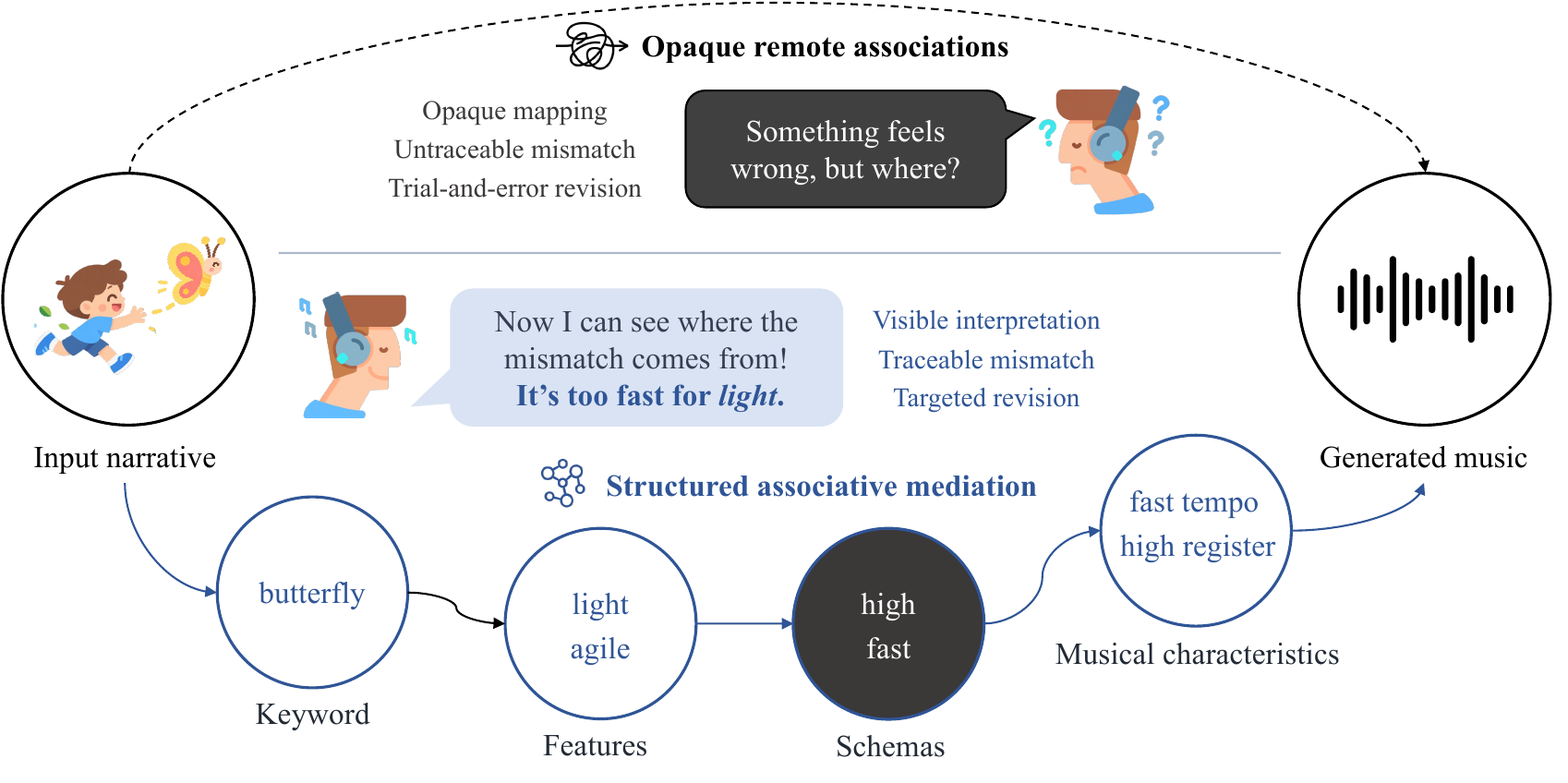}
  \caption*{\textbf{From opaque remote associations to structured associative mediation.} The teaser contrasts how users translate narrative intents into music through an opaque associative pathway versus an explicit, schema-mediated process, highlighting differences in the visibility, negotiability, and perceived control of the creative interaction.}
  \Description{Comparison of opaque versus structured narrative-to-music generation. The top path shows an opaque mapping resulting in untraceable errors and trial-and-error revision. The bottom path illustrates structured associative mediation, decomposing the input into keywords, features, schemas, and musical characteristics to support visible interpretation, traceable mismatch, and targeted revision.}
  \label{fig:teaser}
\end{teaserfigure}

\maketitle

\section{Introduction}

Large-scale music generation models have reached impressive technical maturity. Yet users still often lack a clear sense of control or understanding when interacting with them. While these models can produce high-fidelity and diverse musical excerpts, users' ideas and needs are not always adequately expressed, understood, or reflected in the generated outcomes\cite{ref1}. When outputs diverge from expectations, effective intervention remains difficult, especially problematic for non-expert creators\cite{ref2,ref3}.

Researchers in music computing have sought to address this challenge through advances in data annotation, model architectures, and system design \cite{ref4,ref5,ref6}. We conceptualize an important source of this difficulty as \textbf{Associative Divergence}, which can't be fully addressed by computing methods. Music is an abstract art form without a clear and universally shared system of signification\cite{ref7}. Both musical creation and perception rely on associations evoked by sound. For example, interpreting melodic changes is analogized to fluctuations in emotion; rhythmic density is analogized to the accumulation of energy. Such associations do not follow fixed pathways, they are shaped by multiple factors, including prior experience, memory, and emotional state \cite{ref8,ref9,ref10,ref11}.

Our preliminary study found that, in human-AI music co-creation, users tend to express their ideas through narratives or storytelling. They often lack the ability to articulate a specific musical design, instead hoping that AI can `empathize' with them and infer their intentions. For images or text, people can easily associate perceptual content with narrative themes. For example, one can describe the \emph{Mona Lisa} as depicting a smiling woman. With music, however, it is difficult to identify, without auxiliary information, which particular passage of \emph{Moonlight Sonata} `represents' moonlight. This asymmetry reveals a crucial distinction: translating narratives into musical representations involves a \textbf{remote association}.

This idea is grounded in the associative hierarchy theory from psychology \cite{ref12}. Previous research has shown that, compared with close and accessible associations, forming remote associations is a scarce creative capacity \cite{ref13,ref14}. Music creation precisely requires this kind of cross-modal remote association between narrative concepts and sound. The instability and individual variability of this process contribute to communication difficulties in music creation. When users attempt to express creative intents, such associations leave them unable to exert meaningful control. After AI generates the music, users may also struggle to understand why particular sounds correspond to their descriptions, resulting in lower acceptance of the outcomes \cite{ref1,ref2,ref15}. As prior work has suggested, the use of intelligent creative tools can sometimes reduce deep reflection, and vague intents often lead to equally vague outcomes \cite{ref16}.

Accordingly, this study explores whether decomposing remote associations into shorter segments can enhance creative expression and understanding. We introduce the psychological concept of \textbf{schema}, the basic cognitive structures that people use to understand the world. Schemas are abstract generalizations of recurring perceptual experiences, untethered to any particular modality \cite{ref17,ref18,ref19}. Scholars in music cognition have empirically examined schemas that can be evoked by music, including fundamental perceptual dimensions such as high--low, fast--slow, and strong--weak. However, their potential to support actual creative practices has not been sufficiently explored \cite{ref10,ref20}. We propose that schemas can function as associative mediations, providing users with a visible and actionable cognitive bridge between narrative ideas and musical design.

Building on this premise, we propose a \textbf{bidirectional schema-mediated framework} for human--AI music co-creation. The forward pathway, \emph{narrative → schema → music}, helps users decompose remote associations and progressively translate subjective narratives into musical features. The reverse pathway, \emph{music → schema → narrative}, provides structured interpretations of generated outcomes, helping users understand why the music takes a particular form. Our framework is a form of cognitive scaffolding. Rather than constraining users' associations, it indicates the creative cues that should be identified at each stage, guiding users from vague intents toward concrete musical designs through a process of self-discovery.

Our user studies demonstrate that the framework significantly improves users' perceived control and understanding, and identify a promising approach to creative intent alignment. By externalizing and decomposing associative chains, the framework enables users to locate associative divergence and to diagnose sources of communication breakdown. The framework may also provide a basis for identifying shared cross-modal associative patterns, with potential implications for human--AI co-creation in music and other forms of abstract art. The main contributions of this work are as follows:

\FloatBarrier
\begin{itemize}
\item
  We identify \textbf{Associative Divergence} as a core communication problem in human--AI music co-creation. We argue that the remote associations between free-form narratives and actionable musical features contribute to users' lack of control and understanding.
\item
  We propose and empirically evaluate a \textbf{validated cognitive framework} that translates the psychological construct of schema into a mediation to decompose and externalize the cross-modal association in human--AI music co-creation.
\item
  We evaluate the framework through two user studies (N = 188, N = 60). Results show that it enhances creative expression and output understanding, and uncovers how mediation structures enable deeper ideation and shared understanding in human--AI co-creation.
\end{itemize}
\FloatBarrier

\section{Related Work}

\subsection{Musical Schemas and Cross-Modal Associations}

Music is an abstract temporal medium. Through the organization of pitch, rhythm, dynamics, timbre, harmony, and texture, it can evoke associations of spatial, kinetic, and affective qualities such as \textit{rising--falling} and \textit{approaching--receding}. Musical understanding doesn't occur solely within the auditory modality; it often involves organizing sound into cross-modal structures that can be bodily experienced and verbally described.

Constructivism constitutes an important tradition in cognitive psychology, and one of its central concepts, the \textbf{schema}, refers to a basic model through which people understand the world, representing an abstract generalization of a class of perceptual experiences \cite{ref18,ref19}. Schemas can be understood as frameworks or organizing structures of mental activity, with related concepts taking on different emphases across disciplines. For example, in cognitive linguistics, \textbf{image schemas} refer to embodied, foundational cognitive structures \cite{ref8,ref9}. Such schemas can be extended and updated within different structures of background knowledge \cite{ref23}. In the domain of music cognition, Walker, drawing on metaphors in performance, dance, and musical discourse, argued that musical meaning can be understood as a form of structurally organized cognition of movement \cite{ref24}. Building on this perspective, we define the \textbf{musical schemas} discussed in this work as shared perceptual experiences that mediate between narrative and music. A small set of core schemas may originate from fundamental embodied experiences and subsequently be extended and combined into more diverse repertoires.

Schemas actively mediate cross-modal associations, relating perceptual attributes from one modality with another \cite{ref25,ref26}. Research in music psychology has shown that parameters such as melodic contour, dynamics, and tempo can influence listeners' imagined states of motion \cite{ref10}, while pitch, loudness, timbre, and vibrato can affect judgments of tactile qualities such as weight, softness, hardness, and roughness \cite{ref20}. A cross-cultural study further showed that music and bodily movement can express similar emotions through shared dynamic characteristics, including speed, regularity, direction, and magnitude \cite{ref11}. Collectively, these studies demonstrate that reproducible correspondences exist between musical and non-musical experiences. However, such relationships are better understood as many-to-many associative networks than as mechanically one-to-one mappings.

Musical associations are mediated by emotion, expectation, and memory, and are therefore inherently variable \cite{ref27}. For instance, correspondences between music and colors or visual artworks depend strongly on shared affective evaluations \cite{ref28,ref29}. Listeners also organize sound, movement, narrative, and temporal structure into cross-modal expectations, which are influenced by structural regularity, cultural experience, and episodic memory \cite{ref30,ref31}. Therefore, schemas should not be regarded as static labels of specific sounds, but rather as organizing mechanisms that unfold over time, participate in prediction, and are modulated by experience. Studies of individual differences suggest that listeners' mood, personality, cognitive style, and musical training can influence their judgments of musical emotion and preference \cite{ref32,ref33,ref34}. Cross-modal associations in music can be characterized as \textbf{statistically shared, individually variable, and contextually revisable}. This helps explain why the same piece of music may elicit different emotional, narrative, and bodily associations across listeners.

Collectively, prior studies have explained how musical associations emerge, demonstrated their convergence and recurrence at the population level, and identified the mediating roles of individual characteristics. These complex and variable associative pathways pose a distinctive design challenge for human--AI music co-creation: systems can neither assume that such associations are stable nor leave users to negotiate them entirely. Therefore, effective creative communication requires moving beyond the pursuit of supposedly putative correct parameter mappings. \textbf{Instead, it should support users in understanding, articulating, and revising their own associative processes.}

\subsection{Controllability and Interpretability in AI Music Generation}

Automatic music generation has evolved from rule-based models to deep neural networks. Generative systems now support a broad range of tasks, including composing, arrangement, style imitation, and polyphonic generation \cite{ref4,ref5,ref6,ref35}. Text, attribute labels, and multimodal conditions have increasingly become important entry points of interaction, enabling users to obtain creative results with less effort \cite{ref36}. However, advances in generative capability don't automatically translate into better co-creative experiences. Whereas generative models primarily aim to produce music that satisfies certain quality or conditioning requirements, co-creative systems must support creative ideation and iterative refinement within specific contexts.

Natural language provides an accessible interface for music generation, but it also exposes an interpretive gap between linguistic concepts and musical control. Wang et al. observed that expertise influences users' articulation of the same musical goal, varying in vocabulary and abstraction \cite{ref3}. Metaphorical descriptions, such as ``a suppressed emotion being released,'' may involve multiple musical attributes, including timbre, harmony, and rhythmic density. When a system returns only a complete musical outcome, users may feel difficult to identify where a mismatch originates.

To mitigate this problem, some research has explored more intuitive forms of input and interpretable representations to lower the barrier to control. \emph{Drawlody} allows users to generate melodies by drawing melodic contours, using simplified representation to improve comprehensibility for non-expert users \cite{ref37}. Another line of work structures the generation process into musical hierarchies. \emph{SchenkComposer} organizes phrase structure, harmonic rhythm, harmonic progression, foreground rhythm, and melody into a sequence of stages, allowing users to participate progressively in a generation process grounded in music theory \cite{ref38}. Research on lyrics-to-melody generation has similarly explored local relation modeling and attribute recommendation to help users understand how lyrical content influences melodic generation \cite{ref39,ref40}. These studies suggest that introducing structured intermediate information (contours, attributes and hierarchical representations) between user input and the final music can provide users with meaningful points of intervention, rather than requiring them to interact directly with black-box outputs.

However, these intermediate representations are primarily derived from established music theory or from the internal structure of the generation pipeline. They can't explain how stories and subjective experiences are translated into musical characteristics. For non-experts, concepts such as melodic contour or pitch register are treated as technical vocabulary rather than actionable creative units. Harmon's narrative-inspired music generation system demonstrates that natural language can serve as a conditioning mechanism for music generation \cite{ref41}. But directly mapping semantics to fixed parameters restricts users' ability to express personal and underrepresented interpretations.

We characterize a central gap in human--AI music co-creation as a remote association. Users often express their intents through stories and metaphors, whereas systems require specialized prompts or musical parameters. Between narrative semantics and actionable musical features, there remains a lack of intermediate representations that users and AI can jointly understand and use. Accordingly, this work investigates how to establish a \textbf{visible, audible, and revisable mediation structure} between narrative intents and musical parameters.

\subsection{Interaction Design for Human--AI Music Co-Creation}

Research on human--AI music co-creation has explored a range of interaction mechanisms, including embodied interaction, visualization, parameter editing, and multimodal input \cite{ref2,ref15,ref42,ref43}. These approaches have progressively shifted AI from a one-shot content generator toward an interactive creative partner. HCI research has emphasized that effective co-creation depends not only on output quality, but also on users' perceived control, ownership, self-efficacy, and ability to negotiate with AI \cite{ref44}.

Some work focuses on making musical control more accessible through embodied and visual interaction. Systems such as \emph{Touching Wires}\cite{ref45}, \emph{Co-Performing Music with AI}\cite{ref46}, and \emph{Sonified Body}\cite{ref47} establish real-time feedback loops through bodily movement, gesture, tangible interfaces, or haptic interaction. Other systems provide visual or parameterized controls. Louie et al., for example, introduced part-level constraints, similarity controls, semantic sliders, and candidate auditioning to support novice users in steering generative models \cite{ref2}, while \emph{Drawlody} used melodic contour as an interpretable visual mediation for expressing melodic direction \cite{ref37}. These approaches lower the operational barrier to controlling music generation. However, users are still required to determine how such gestures, contours, or parameters relate to their own stories, emotions, and creative intents.

A second line of research allow users to guide music generation through cross-modal materials, such as text, images, or musical examples\cite{ref68,ref50,ref51}. Chung et al.~organized image references and musical motifs through schemas, associations, and cross-domain mappings, providing an early precedent for schema-mediated interaction \cite{ref48}. More recent work has examined prompt-based, preset-based, and motif-based paradigms, showing that users with different musical backgrounds engage with these forms of input in different ways \cite{ref49}. These studies demonstrate that cross-modal inputs can make musical creation more accessible to non-experts. However, such translations are one-directional and operate at the level of concrete musical parameters, leaving the user's higher-level narrative intent outside a negotiable loop.

Broader research on human--AI co-creation similarly suggests that effective collaboration requires iterative negotiation, visible roles, and opportunities for users to inspect and correct AI contributions \cite{ref15,ref42,ref43,ref52,ref53}. However, existing music co-creation interfaces primarily make operations visible (such as parameters, gestures, prompts, or alternatives) while the associative process that connects users' intents to those operations often remains implicit.

We address this question by externalizing the path from \emph{narrative themes → perceptual attributes → schemas → musical features}. Recent work has also begun to examine authority relations in human-AI music co-creation; for example, Hu et al. proposed a framework built on contestability, agency, and plurality \cite{ref67}. While we share this concern for creator agency, our focus differs: rather than asking who holds interpretive authority, we ask how the association chain between narrative and musical structure can be made visible and negotiable. Our framework complements existing interfaces by providing a shared intermediate structure for organizing creative intents.

\section{Analogous Study: Insights from Human Artist--Client Collaboration}

For non-expert users, AI can provide artistic skills that exceed their own. This form of collaboration resembles private art commissions. In such settings, clients contribute their own stories and ideas as the creative theme, while artists interpret the theme and use their professional expertise to realize it. So we use the human artist--client collaboration as a \textbf{structural analogy} to human--AI co-creation. From the perspectives of clients and artists, we examine how creative intents are expressed, interpreted, and negotiated in the collaboration, and derive design strategies that may be transferable to human--AI co-creation.

\subsection{User-Side Finding: A Narrative Preference in Creative Expression}

We conducted a questionnaire study to investigate users' motivations for music creation, willingness to participate in different stages of the creative process, and preferred ways of expressing creative ideas. The survey was distributed through the public data collection platform Credamo (\url{https://www.credamo.com/}). We obtained 194 valid responses. Participants ranged in age from 19 to 60 years (M = 30.8, SD = 8.76), including 83 men and 111 women. 37 participants had received formal musical training, while 157 reported amateur-level musical experience. The main findings are summarized below.

\begin{figure}[H]
  \centering
  \includegraphics[width=\textwidth]{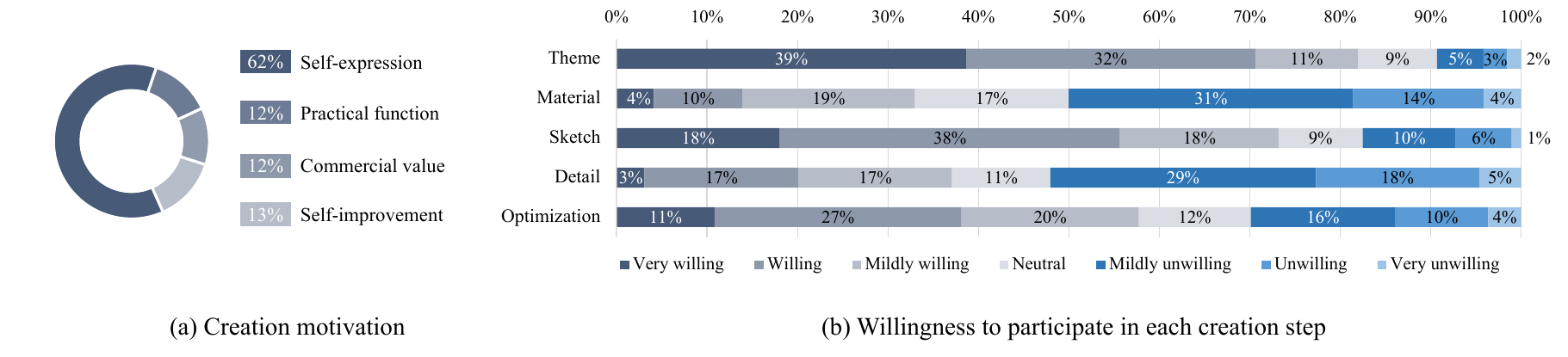}
  \caption{Music creation motivations and willingness to participate: (a) distribution of motivations for music creation; (b) willingness to participate across different stages of the creative process.}
  \Description{Two-part survey chart displaying creation motivation (a) and step-by-step participation willingness (b). A donut chart shows Self-expression as the leading motivation (62\%) over practical function, commercial value, and self-improvement. A horizontal stacked bar chart shows participants express the highest willingness to participate in Theme and Sketch stages, while showing higher reluctance towards Material, Detail, and Optimization steps.}
  \label{fig:motivation}
\end{figure}

\textbf{(1) Creative Motivation and Willingness to Participate.} The results in Fig.~\ref{fig:motivation}a show that self-expression was a primary creative motivation for 62\% of participants. Participants expressed greater willingness to engage in stages that directly reflected their creative intents, such as theme defining and melodic shaping. In contrast, for stages requiring greater domain expertise, such as material collecting, refinement, and optimization, participants were more inclined to seek assistance from professional creators (see Fig.~\ref{fig:motivation}b). 

\textbf{(2) Creative Expression Patterns.} In an open-ended question in the survey, we asked participants to provide a description of a creative idea that they would use to communicate with an AI system. After excluding 13 responses whose meanings were unclear, we identified five categories of information that frequently appeared in participants' descriptions: (1) story, (2) emotion, (3) viewpoint, (4) musical parameters, and (5) usage. As shown in Table~\ref{tab:expression-types}, participants tended to express their musical ideas through stories, emotions, and concrete scenarios, while directly specifying musical parameters was comparatively less common (combinations with frequency less than 5 are not displayed).

These results suggest that \textbf{users know what to express, but not how to express it musically}. What they need is not simply direct access to parameter controls, but a translation process that can bridge linguistic creative intents and professional musical realization.

\begin{table}[htbp]
\centering
\small
\renewcommand{\arraystretch}{1.2}
\caption{Frequency of different types of information appearing in descriptions of musical creative ideas.}
\label{tab:expression-types}
\begin{tabular}{lcp{9.5cm}}
\toprule
\textbf{Type} & \textbf{Frequency} & \textbf{Case (user code)} \\
\midrule
\multirow{4}{*}{Story} 
  & \multirow{4}{*}{61} & A whale falls and all things flourish. (U44) \\
  &                     & The journey of pursuing my dreams. (U158) \\
  &                     & People break free from the shackles of society and run freely across the land of Utopia, singing songs of freedom. (U160) \\
\midrule
\multirow{2}{*}{Usage} 
  & \multirow{2}{*}{8}  & I hope the music can promote Chinese culture. (U36) \\
  &                     & Commemorate the heroes of a special era. (U59) \\
\midrule
\multirow{2}{*}{Parameter} 
  & \multirow{2}{*}{13} & Make some adaptations based on DJ dance music. (U24) \\
  &                     & A soothing rhythm paired with a group chorus. (U94) \\
\midrule
\multirow{2}{*}{Emotion} 
  & \multirow{2}{*}{20} & Express a reluctance to let go of youth. (U28) \\
  &                     & It should be passionate and inspiring. (U64) \\
\midrule
View 
  & 6                   & Youth is limitless, and youth is carefree. (U159) \\
\midrule
Story \& Usage 
  & 6                   & Showcase the lives of left-behind children to raise awareness about this issue. (U176) \\
\midrule
\multirow{2}{*}{Story \& Emotion }  
  & \multirow{2}{*}{27}                 & I moved from the noisy, bustling, light-polluted city to the quiet, dark countryside mountains. Gazing at the starry sky, I expressed deep longing for peace. (U37) \\
\midrule
Usage \& Parameter 
  & 5        & I want to write a rhythmic piece as background music for exercise. (U12) \\
\midrule
\multirow{2}{*}{Usage \& Emotion}   
  & \multirow{2}{*}{6}                  & It should convey enthusiasm and positivity, uplifting listeners and inspiring them to be more optimistic. (U115) \\
\midrule
\multirow{2}{*}{Parameter \& Emotion}  
  & \multirow{2}{*}{9}                  & Gather sounds from the natural world, then incorporate them into the piece. Choose suitable tones to create a calm and peaceful atmosphere. (U66) \\
\bottomrule
\end{tabular}
\end{table}

\subsection{Artist-Side Findings: Guiding, Translating, and Aligning}

We conducted semi-structured interviews with 17 practitioners experienced in commissioned art creation (A1--A17). Participants were recruited through an open call posted in art-related communities on Douban (\href{https://m.douban.com/}{https://m.douban.com}). They had an average of 4.3 years of professional experience (range: 0.5 to 13 years), and all provided relevant records verifying their work experience. Each interview lasted approximately 30 minutes and was conducted remotely via an online voice call. The interviews focused on creative communication processes, challenges in client collaboration, and aligning strategies. Our analysis revealed three core strategies artists use to help clients articulate and realize creative intent: \textbf{guidance, translation, and alignment}.

\textbf{(1) Guidance.} From the artists' perspective, a major barrier was that clients often lacked a clear understanding of their own intent, risking the omission of important creative information (A07). As one artist explained, ``I would still guide them rather than reject their ideas'' (A03). Some artists developed standardized workflows that specified the information clients should consider and provide at different stages of the collaboration (A01, A06, A12). Such processes helped clients complete their ideas, clarify the acceptable scope of revisions (A10), and develop more realistic expectations about the outcomes (A04).

\textbf{(2) Translation.} Differences in domain expertise constituted another communication difficulty (A12). Some artists attempted to teach clients relevant professional concepts to elicit more precise information about their creative intents (A11). However, as one artist noted, ``If every client has to be trained first, the communication cost becomes too high'' (A13). Consequently, some artists expressed a desire for ``a tool that could guide users in translating their own ideas'' (A17). They also mentioned a specialized role, \textbf{the liaison}, who serves as an mediation between clients and creators and is responsible for translating and clarifying clients' requirements (A15, A08).

\textbf{(3) Alignment.} Most artists described an iterative process in which creation and clarification occurred in parallel, with the work adjusted as clients' intents became clearer (A02, A05, A16). As one participant stated, ``Some clients like us to show progress and ask for feedback frequently, because it makes them feel like participating in the creation'' (A14). In addition, some artists maintained libraries of commonly used demos. ``Based on the demos, clients might tell us that they want a more intense emotion, a similar timbre, or a combination of two styles'' (A15). These examples provided a shared reference through which clients and artists could negotiate and reach a consensus.

\subsection{Design Insight: A Flexible Mediation Strategy}

Through the analogous study above, we found that effective communication relies on a \textbf{flexible mediation strategy} that creates a gradual transition between users' free-form expressions and artists' professional practices. Based on this finding, we argue that the interaction process of human--AI music co-creation should be redesigned around the following three principles.

\textbf{(1) Structured Guidance.} The standardized workflows used by artists to guide clients can be a form of structured constraint. Unlike fixed parameter-mapping rules, such a process should function as staged cognitive guidance. It should prompt users to consider and clarify particular types of information at different stages, thereby establishing an interpretable and decomposable chain of creative reasoning rather than restricting users' freedom of choice.

\textbf{(2) Narrative Translation.} Narrative serves as a natural starting point for creative expression. Interaction design for human--AI co-creation should respect this practice rather than requiring users to adopt parameterized or domain-specific language. More importantly, translating narratives into professional representations should be a process of alignment, rather than an opaque internal operation of AI models. Interaction design should not exclude users from the translation process but provide effective scaffolding that enables them to participate and negotiate.

\textbf{(3) Cognitive Alignment.} The generated outcomes should be provided with appropriate interpretations of how they were produced. Such interpretations can help users understand how their creative expressions have been realized, while also enabling them to identify divergence and determine appropriate revision strategies.

\section{Method: The Bidirectional Schema-Mediated Cognitive Framework}

Section 3 shows that creative collaboration between non-expert users and professional creators is fundamentally a continuous process of guidance, translation, and alignment. This suggests that what human--AI music co-creation requires is not simply more controllable parameters, but a flexible mediation that can accommodate natural narratives, organize intermediate creative reasoning, and support inspection and revision.

Drawing on prior research on cross-modal association and music cognition \cite{ref18,ref19}, we consider \textbf{schemas} a promising mediating structure. Schemas can establish interpretable and visible associative between cross-modal perceptual features. We operationalize the flexible mediation strategy proposed in Section 3 as a \textbf{bidirectional schema-mediated framework}. The framework consists of two complementary pathways: \emph{narrative → schema → music}, which supports creative expression, and \emph{music → schema → narrative}, which supports users in understanding generated outcomes.

\subsection{Musical Feature--Schema Association Map}

The shared schemas discussed in this work do not refer to fixed psychological structures that are universally shared by all individuals. They denote a set of schemas that have shown relatively stable associations in prior studies of music perception and can serve as a basis for cross-modal expression. Compared with highly individualized schemas that depend strongly on personal experience, these schemas have a broader shared basis\cite{ref17,ref54}.

Through a series of systematic perceptual experiments, Zhou Haihong identified five shared schemas that can be evoked by musical features: \textbf{high--low, strong--weak, fast--slow, soft--hard, and tense--harmonious} \cite{ref55}. The experiments showed that these schemas exhibit relatively direct and stable relationships with corresponding musical features. For example, \emph{high--low} primarily corresponds to pitch, \emph{fast--slow} to tempo, \emph{strong--weak} to dynamics, \emph{soft--hard} to timbre and articulation, and \emph{tense--harmonious} to harmony and degree of dissonance.

\begin{figure}[htb]
  \centering
  \includegraphics[width=\textwidth]{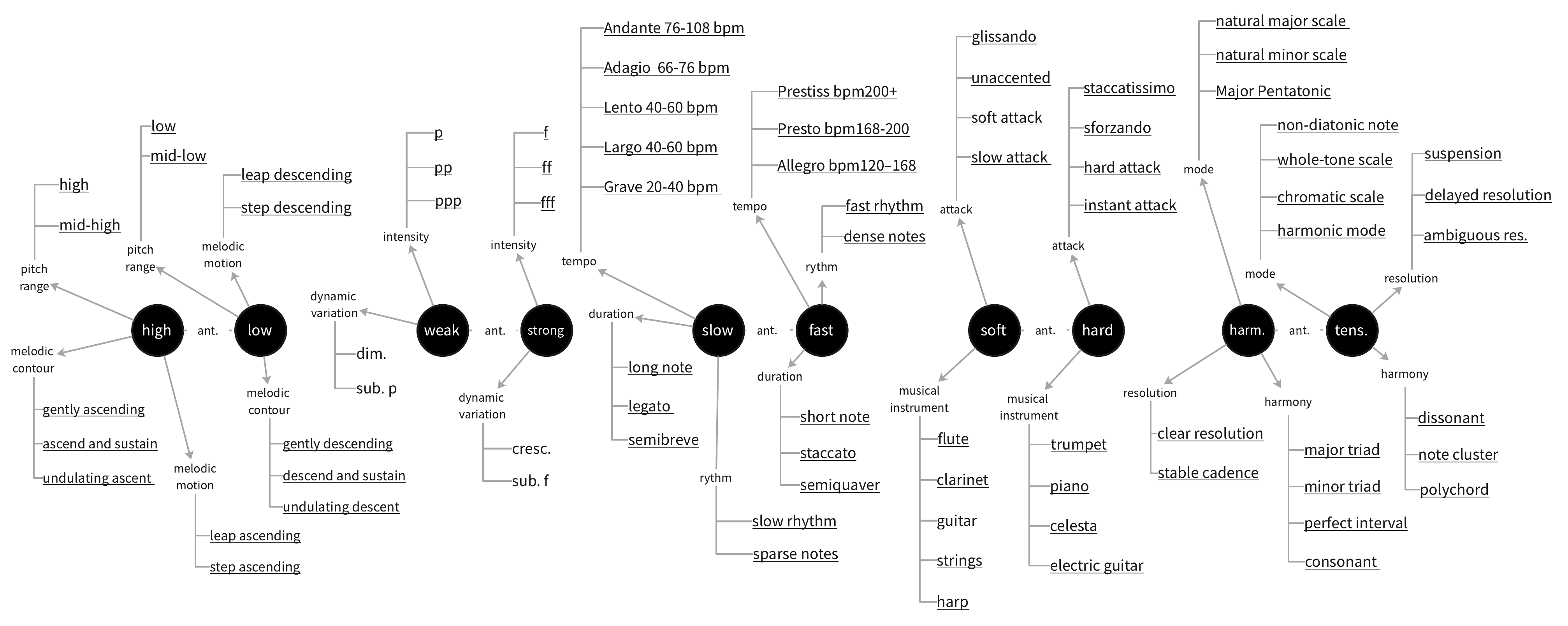}
  \caption{Music feature-schema association map.}
  \Description{A hierarchical mind map of musical feature schemas organized into six bipolar pairs: high vs. low, weak vs. strong, slow vs. fast, soft vs. hard, and harmonious vs. tense. Each pair branches outward to define specific musical characteristics across dimensions including pitch range, melodic contour, intensity dynamics, tempo, duration, attack, instrumentation, mode, resolution, and harmonic consonance/dissonance.}
  \label{fig:feature-schema}
\end{figure}

Based on findings from prior empirical studies, we categorized the musical features that can evoke schema-based associations \cite{ref10,ref20,ref55} and constructed a \textbf{musical feature--schema association map} (Fig.\ref{fig:feature-schema}). Importantly, this map is not intended to enumerate all professional musical descriptors. It is designed to help users understand the relationships between musical features and schemas, while providing reasonable options and examples. The map also allows users to introduce additional or analogous musical features according to the creative context.

\subsection{Narrative--Schema Association Map}

\begin{figure}[htb]
  \centering
  \includegraphics[width=\textwidth]{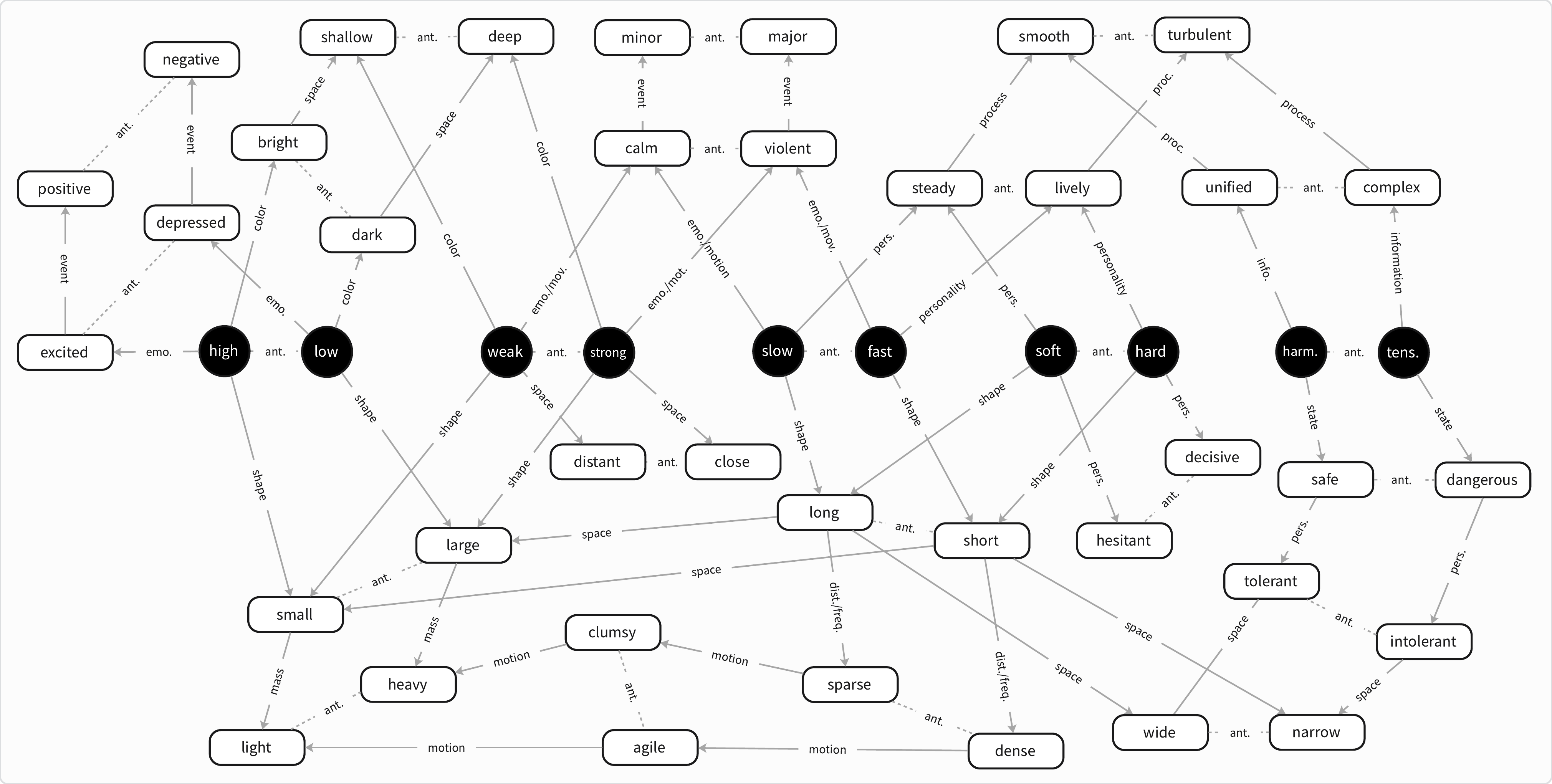}
  \caption{Narrative-schema association map.}
  \Description{Cross-modal semantic network expanding five core musical schema pairs (high/low, weak/strong, slow/fast, soft/hard, harmonious/tense) into multi-dimensional perceptual concepts. Labeled edges link musical schemas to opposing cross-domain antonym pairs representing emotion, space, mass, motion, color, personality, process, and state attributes.}
  \label{fig:narrative-schema}
\end{figure}

Simple combinations of the five core schemas are insufficient to accommodate the richness of narrative intents. Prior research on cross-modal association suggests that core schemas can serve as mediation structures linking to a broader range of perceptions \cite{ref27}. Specifically, the five core schemas can be associated with analogous schemas across visual \cite{ref28,ref55}, tactile \cite{ref20,ref55}, affective \cite{ref11,ref28,ref55}, spatial and motion-related \cite{ref10,ref11} dimensions. For example, \emph{high--low} may extend to associations such as \emph{bright--dim}, \emph{light--heavy}, and \emph{excited--calm}; \emph{fast--slow} may extend to affective dimensions such as \emph{urgent--relaxed} and \emph{restless--composed}; and \emph{strong--weak} may extend to descriptions of motion states such as \emph{intense--gentle}. Based on these findings, we expanded the narrative--schema association map (Fig.\ref{fig:narrative-schema}).

In this map, the core schemas (shown as black circles) function as mediation nodes that can both connect to the musical-feature layer and extend toward narrative. Combined with the music feature--schema association map introduced in Section 4.1, the two maps form a complete associative pathway from \textbf{narrative → schema → musical features}. Starting from any theme or narrative, users can navigate through the map to identify perceptual schemas that resonate with their intents and then locate actionable musical features.

The associations in the map are not intended to be universally correct. The map just presents a range of plausible associative pathways intended to stimulate users' own reflection and interpretation. This is also why we adopt a graph-based representation: extended schemas and their relationships can be dynamically added, revised, or reorganized according to different associative habits, preferences, and creative contexts. Through continued use, the map can become progressively richer and more personalized.

\subsection{Bidirectional Mediated Interactive Process}

\begin{figure}[H]
  \centering
  \includegraphics[width=0.8\textwidth]{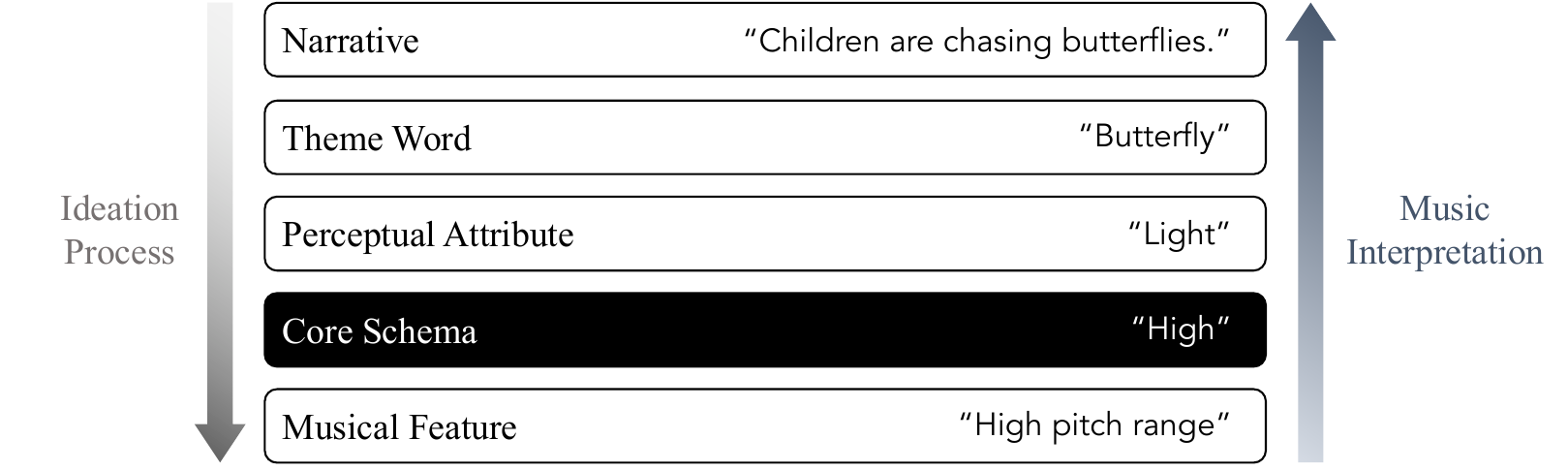}
  \caption{Schema-guided music ideation workflow and music interpretation strategy based on bidirectional mediation.}
  \Description{A five-step linear workflow showing how narrative input ("Children are chasing butterflies") translates through Ideation Process ("Butterfly" to perceptual attribute "Light") and Music Interpretation (core schema "High" to musical feature "High pitch range").}
  \label{fig:workflow}
\end{figure}

Based on the association maps described above, we designed a schema-guided music ideation workflow and a music interpretation strategy, as shown in Fig.~\ref{fig:workflow}. In the forward pathway, users begin with free-form narratives and progressively derive musical features through schema-based associations. In the reverse pathway, the framework links generated outcomes back to the corresponding musical features and schemas, enabling users to understand how their creative intents have been realized and to identify potential communication mismatches. The maps introduced in the previous two sections provide a structured knowledge reference and cognitive framework for human--AI music co-creation. The ideation workflow and interpretation strategy presented in this section, in turn, provide a more explicit operational logic for \textbf{translation} and \textbf{alignment}.

\subsubsection{Staged Music Ideation Process}

The music ideation process consists of five core steps that guide users in clarifying their creative intents and translating narratives into professional musical features:

\textbf{(1) Narrative Elicitation.} Users express their initial ideas in the form that feels most natural to them, for instance, as a short story, a set of keywords, or a scene description (e.g., ``a child is chasing a butterfly'').

\textbf{(2) Theme Word Extraction.} As an abstract medium, music can't explicitly represent highly complex narrative information. The narrative needs to be distilled into one or two \textbf{theme words}. These theme words may refer to objects, actions, emotions, or other relatively concrete elements. For example, from the narrative ``a child is chasing a butterfly,'' the words ``butterfly'' and ``joy'' may be extracted as the \textbf{theme words}.

\textbf{(3) Perceptual Attribute Decomposition.} The attributes of each theme word are analyzed so that they can be associated with schemas in the map. This step guides users in abstracting the concrete objects or concepts. For example, the theme ``butterfly'' may be decomposed into attributes such as ``delicate'' and ``agile''  while ``joy'' may be decomposed into ``positive'' and ``relaxed.''

\textbf{(4) Schema Identification.} Using the narrative--schema association map in Section 4.2, each attribute can be associated with the perceived closest core schema. For instance, ``delicate'' may correspond to \emph{weak}, ``agile'' to \emph{fast}, ``positive'' to \emph{high}, and ``relaxed'' to \emph{harmonious}. At this stage, users are encouraged to avoid selecting mutually opposing schema values, such as \emph{strong} and \emph{weak} simultaneously. If such conflicts can't be avoided, they can be resolved or balanced during the subsequent musical-feature design stage.

\textbf{(5) Musical Feature Design.} Using the music feature--schema association map in Section 4.1, users select concrete combinations of musical features based on the identified core schemas. For example, given the four schemas \emph{weak}, \emph{fast}, \emph{high}, and \emph{harmonious}, the user may specify ``diminuendo,'' ``fast rhythm,'' ``higher pitch range,'' and ``harmonic mode'' as explicit musical instructions. Contradictory feature combinations, such as specifying both low and high dynamics, should be avoided.

This process decomposes the single remote association into five short and inspectable stages. At each stage, users can examine, adjust, or revise their judgments, progressively transforming vague narrative intents into actionable musical control instructions.

\subsubsection{Schema-Based Music interpretation Strategy}

The reverse application of the framework provides a structured strategy for explaining generated musical outcomes. Unlike the staged interaction in the ideation process, the interpretation process doesn't require users to participate step by step. Instead, the complete associative pathway is externalized and presented to users through a visual representation.

The interpretation pathway begins with musical features and traces them, through schemas, back to perceptual attributes and narrative themes. Specifically, the framework organizes structured information to explain: which concrete musical features were used in the generated music, such as tempo, pitch, and dynamics; which core schemas these parameters correspond to, such as \emph{fast} or \emph{high}; how these schemas relate to perceptual attributes associated with the theme words, such as \emph{agile} or \emph{positive}; and ultimately how these associations connect back to the narrative themes originally expressed by the user.

This interpretation strategy helps users identify potential perceptual mismatches. For example, for the theme word ``butterfly,'' a fast rhythm intended to convey ``agility'' may be perceived by the user as excessively fast, thereby weakening the intended sense of ``lightness.'' The user can then locate the source of this divergence at the schema level, recognizing that the balance between the \emph{fast} and \emph{lightness} needs adjustment, and provide clear feedback for revision.

\section{User Study: Exploring Expression and Understanding}
To evaluate the effectiveness of the bidirectional schema-mediated framework, we designed and conducted two user studies. Study 1 focused on how the forward pathway influences users' creative expression and perceived control. Study 2 examined how the reverse pathway affects users' understanding and evaluation of generated outcomes. Together, the two studies address the following research questions:

\FloatBarrier
\begin{itemize}
\item
  How does the framework influence users' musical creative expression across different narrative themes?
\item
  How do users' associative pathways between narrative ideas and music features?
\item
  Does the framework-guided process enhance users' perceived control during ideation?
\item
  Does the interpretation provided by the framework improve users' understanding of generated outcomes?
\end{itemize}
\FloatBarrier

\subsection{Study 1: Enhancing Creative Expression and Perceived Control}

Study 1 evaluated whether the staged ideation process in Section 4.3.1 could improve the specificity of users' musical creative expression and their perceived sense of control. We adopted a between-subjects design comparing the proposed framework with a conventional musical terminology reference in supporting the ideation process. This study evaluates the effectiveness of the \textbf{cognitive framework}, rather than a system implementation. Participants used the framework to develop their creative ideas without interacting with a live AI model. This setup decoupled the cognitive effects of the framework from other factors, such as UI layout, response latency, and generation quality.

\subsubsection{Participants}

We recruited 188 participants through the public online research platform Credamo (\url{https://www.credamo.com/}), with 94 participants assigned to the experimental condition and 94 to the control condition. All participants were recruited from China. Their mean age was 31.1 years (SD = 9.67), including 85 men and 103 women. All participants were informed of the study procedure, consented to the collection and use of their data, and received monetary compensation for their participation.

\subsubsection{Experimental Design}

\begin{figure}[htb]
  \centering
  \includegraphics[width=\textwidth]{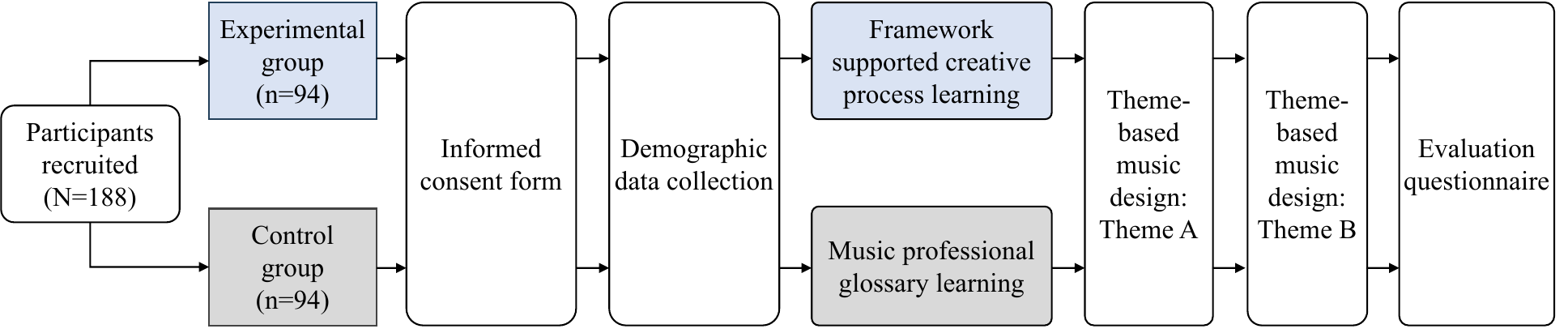}
  \caption{Procedure of Study 1.}
  \Description{Flowchart of the between-subjects study procedure for 188 participants split into experimental and control groups ($n = 94$ each). After consent and demographic collection, the experimental group receives framework-supported creative process learning while the control group receives music glossary learning, followed by theme-based music design (Themes A and B) and an evaluation questionnaire.}
  \label{fig:study1-procedure}
\end{figure}

Participants were randomly assigned to either the experimental or control condition. The procedures for the two conditions are illustrated in Fig.~\ref{fig:study1-procedure}. The study was conducted online and lasted approximately 30 minutes. Before beginning the experiment, all participants were informed of the study purpose and procedure and provided informed consent.

Participants in the experimental group first watched an instructional video introducing the schema framework, including an overview of the schema association maps and a demonstration of the staged ideation process. Participants in the control group were instead provided with a reference sheet of commonly used musical terms together with instructions for their use. The materials explained core concepts including pitch, rhythm, dynamics, timbre, and mode, and provided example expressions and usage scenarios. Participants were given six minutes to study these materials.

After confirming that they understood how to use the assigned support materials, participants in both groups sequentially completed two creative tasks based on Themes A and B. For each theme, participants were asked to design a set of musical features using the guidance provided in their respective condition and then produce a final text prompt. To facilitate subsequent analysis, we standardized several aspects of the task: participants were asked to extract one or two core theme words for each theme, and participants in the experimental condition could select no more than four schemas. The platform automatically recorded ideation time. After completing both tasks, all participants completed a subjective evaluation questionnaire and provided open-ended feedback.

\begin{figure}[bth]
  \centering
  \includegraphics[width=\textwidth]{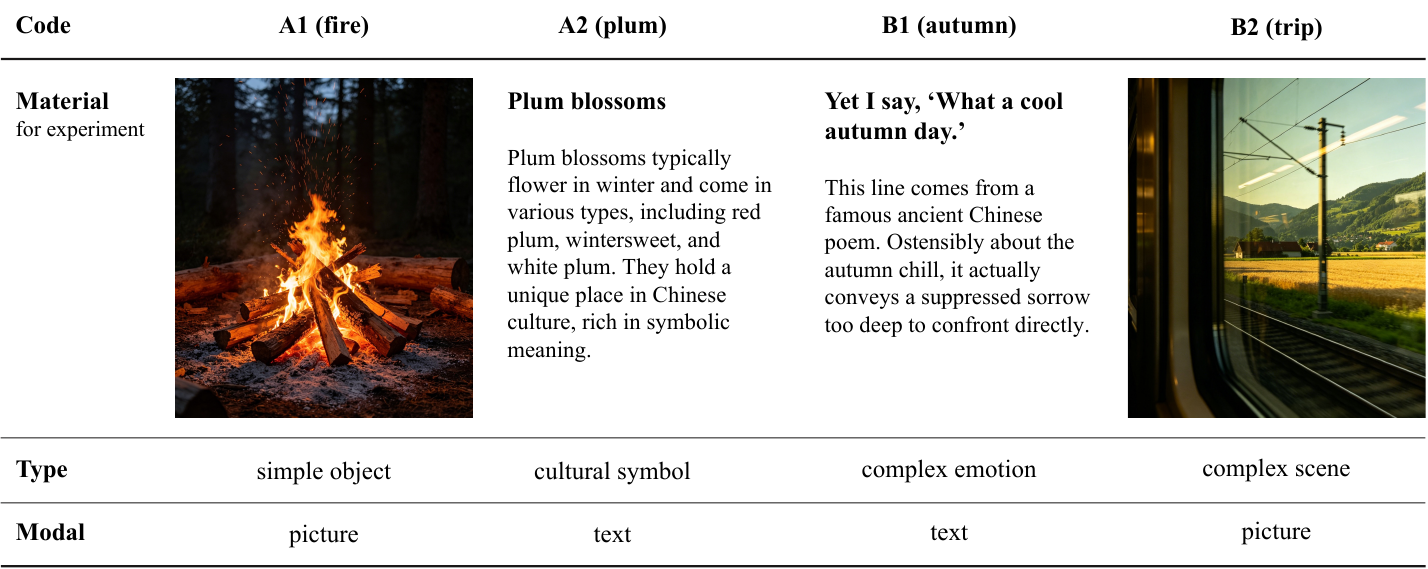}
  \caption{Four creative themes used in Study 1.}
  \Description{Experimental stimuli for four themes categorized by type and modality: A1 ("fire", simple object, picture of a campfire), A2 ("plum", cultural symbol, descriptive text), B1 ("autumn", complex emotion, poetic text), and B2 ("trip", complex scene, picture of scenery viewed from a train window).}
  \label{fig:themes}
\end{figure}

\subsubsection{Creative Themes}

We designed four creative themes to cover different types of creative challenges, as shown in Fig.~\ref{fig:themes}. In both groups, half of the participants were assigned Themes A1 and B1 for the two creative tasks, while the other half were assigned Themes A2 and B2. The four themes were designed to vary in modality, complexity, and degree of cultural specificity, allowing us to examine the applicability of the framework across a broader range of creative contexts.

\Needspace{11\baselineskip}
\subsubsection{Data Collection}

We collected the following three types of data:

\FloatBarrier
\begin{itemize}
\item
  \textbf{Creative Description Texts.} The final descriptions of musical ideas produced by participants for each theme were collected for word-frequency analysis and comparison of expression characteristics.
\item
  \textbf{Ideation Time.} We recorded task completion time to characterize differences in the amount of time participants spent on the ideation process.
\item
  \textbf{Subjective Evaluation Questionnaire.} A seven-point Likert scale was used to measure participants' \textbf{Perceived Effectiveness}, \textbf{Perceived Ease of Use}, and \textbf{Perceived Control Enhancement} with respect to the assigned support tool.
\end{itemize}

\subsection{Study 2: Enhancing the Understanding of Generated Outcomes}

Study 2 investigated whether the schema-based music interpretation strategy described in Section 4.3.2 could improve users' understanding of AI-generated music and their perceived \textbf{Theme Relevance}. We adopted a between-subjects design comparing participants' evaluations of the same musical pieces under conditions with and without  interpretation information.

\subsubsection{Experimental Materials}

From the creative descriptions produced by the experimental group of Study 1, we selected two types of prompts (see Table \ref{tab:musical_features}). \textbf{High-frequency prompts} represented conventional mappings between themes and schemas and reflected associative pathways selected by a larger proportion of participants. \textbf{Low-frequency prompts}, in contrast, represented less conventional theme--schema mappings and more distinctive associative expressions selected by relatively few participants. For each of the four themes, we selected one prompt of each type and submitted them to an AI music generation platform (\url {https://www.sunoai.cn/backgroundMusic}), resulting in eight musical excerpts in total. All excerpts were generated using the same AI music generation model based on their corresponding prompts and were subsequently standardized to maintain consistency in audio quality and duration.

\begin{table}[htbp]
\centering
\small
\caption{Musical Features and Perceptual Attributes Across Themes}
\label{tab:musical_features}
\begin{tabularx}{0.95\textwidth}{l c c X}
\toprule
\textbf{Theme} & \textbf{Regularity} & \textbf{Core Schemas} & \textbf{Musical Features} \\
\midrule
\multirow{4}{*}{A1 (fire)} 
  & \multirow{2}{*}{Conv.}   & \multirow{2}{*}{high, strong}   & (1) High register, expressing the brightness\\
  &                                 &                                & (2) Prominent accents, expressing the heat \\
\cmidrule(lr){2-4}
  & \multirow{2}{*}{Unconv.} & \multirow{2}{*}{fast, tension}  & (1) Fast tempo / allegro, expressing the leaping \\
  &                                 &                                & (2) Dissonant intervals, expressing potential danger \\
\midrule
\multirow{4}{*}{A2 (plum)} 
  & \multirow{2}{*}{Conv.}   & \multirow{2}{*}{hard, high}     & (1) Guqin and guzheng timbres, expressing the resilience\\
  &                                 &                                & (2) Higher register and ascending melody, expressing the nobility\\
\cmidrule(lr){2-4}
  & \multirow{2}{*}{Unconv.} & \multirow{2}{*}{low, slow}      & (1) Lower register and descending melody, expressing the introversion\\
  &                                 &                                & (2) Slow tempo / adagio, expressing the elegance\\
\midrule
\multirow{4}{*}{B1 (autumn)} 
  & \multirow{2}{*}{Conv.}   & \multirow{2}{*}{low, tension}   & (1) Lower register, expressing desolation \\
  &                                 &                                & (2) Some dissonance and consonance interwoven, expressing anxiety \\
\cmidrule(lr){2-4}
  & \multirow{2}{*}{Unconv.} & \multirow{2}{*}{slow, soft}     & (1) Slow tempo / adagio, expressing a gentle breeze \\
  &                                 &                                & (2) String timbres, expressing a comfortable mood \\
\midrule
\multirow{4}{*}{B2 (trip)} 
  & \multirow{2}{*}{Conv.}   & \multirow{2}{*}{harmony, high}  & (1) Consonant melodies and chords, expressing beautiful scenery \\
  &                                 &                                & (2) Higher and brighter register, expressing vastness and freedom \\
\cmidrule(lr){2-4}
  & \multirow{2}{*}{Unconv.} & \multirow{2}{*}{fast, soft}     & (1) Fast tempo / allegro, expressing the motion of the train \\
  &                                 &                                & (2) String timbres, expressing freedom and comfort \\
\bottomrule
\end{tabularx}
\end{table}

\subsubsection{Participants}

We recruited 60 participants through the same online research platform, including 26 men and 34 women. Participants were randomly assigned to the interpretation and no-interpretation conditions, with 30 participants in each group. Their ages ranged from 20 to 56 years (M = 30.5, SD = 8.74), and none had participated in Study 1. All participants were informed of the study procedure, consented to the collection and use of their data, and received monetary compensation for their participation.

\subsubsection{Experimental Design}

Participants were randomly assigned to either the \textbf{interpretation condition} or the \textbf{no-interpretation condition}. In the interpretation condition, participants were presented with the complete associative pathway corresponding to each musical excerpt: \emph{musical features → core schemas → perceptual attributes → theme}. In the no-interpretation condition, participants listened to the same musical excerpts but were provided only with the corresponding theme information. The study was conducted online and lasted approximately 40 minutes. Participants first read the study instructions and provided basic demographic information. They then listened to eight musical excerpts in sequence. Before each excerpt, participants were shown the corresponding theme name or image; participants in the interpretation condition additionally viewed the structured associative pathway. After listening to each excerpt, participants rated its theme relevance and comprehensibility. For three of the eight excerpts, participants were also asked to provide suggestions for improving the generated music after completing their ratings.

\subsubsection{Data Collection}

After listening to each musical excerpt, participants rated the following dimensions on seven-point Likert scales: \textbf{(1) Theme Relevance:} To what extent does the music correspond to its theme? To what extent does the music reflects the perceptual attributes of the theme? \textbf{(2) Comprehensibility:} To what extent do you understand how the creative intent is manifested in the music? To what extent do you understand why the author created this music?

In addition, we collected brief revision suggestions for three designated excerpts from each participant. These responses were used to examine whether explanatory information helped users formulate more structured and actionable feedback for revising generated music.

\section{Analysis and Findings}

\subsection{Effects of the Framework on Users' Creative Expression}

\begin{figure}[htb]
  \centering
  \includegraphics[width=0.9\textwidth]{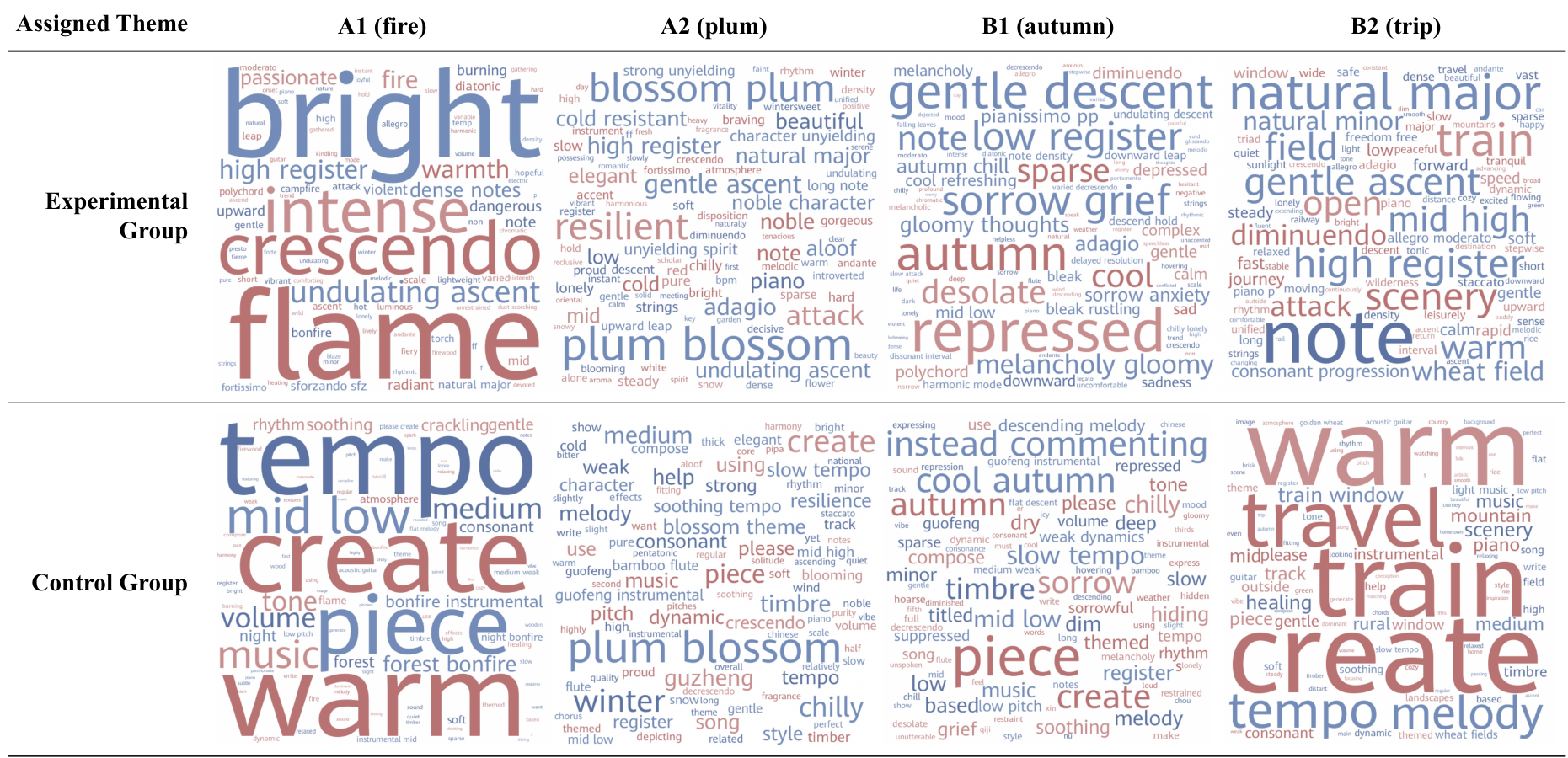}
  \caption{Word-cloud analysis of musical idea descriptions in Study 1.}
  \Description{Word clouds comparing music design rationale keywords between the Experimental Group (top) and Control Group (bottom) across themes A1, A2, B1, and B2. The Experimental Group predominantly features structured musical terms (e.g., register, tempo, contour, dynamics, chord structures), whereas the Control Group relies on informal genre labels, concrete scene objects, and prompt-oriented words (e.g., "guofeng", "wheat fields", "please", "write").}
  \label{fig:wordclouds}
\end{figure}

We used word clouds to visualize the musical idea descriptions produced by the experimental and control groups across different themes (Fig. \ref{fig:wordclouds}). The high-frequency terms reveal clear differences between the two ideation approaches.

First, the experimental group showed greater convergence in their use of concrete musical features. For example, frequently occurring expressions included \emph{crescendo} for the Theme A1 (fire), \emph{gentle descent} for Theme B1 (autumn), and \emph{natural major} and \emph{high register} for Theme B2 (trip). Although the control group also referred to dimensions such as tempo, melody, and volume, they tended to describe them using relatively vague expressions, such as ``a cold timbre'' or ``a powerful melody.''

Second, for the more complex Theme B2 (trip), the control group exhibited more highly shared words. However, these terms tended to summarize the theme rather than interpret and translate it. Words such as \emph{warm}, \emph{travel}, and \emph{train} remained close to the visual materials we provided, but distant from actionable musical features. In contrast, the experimental group produced more diverse yet more executable descriptions.

\begin{figure}[htb]
  \centering
  \includegraphics[width=\textwidth]{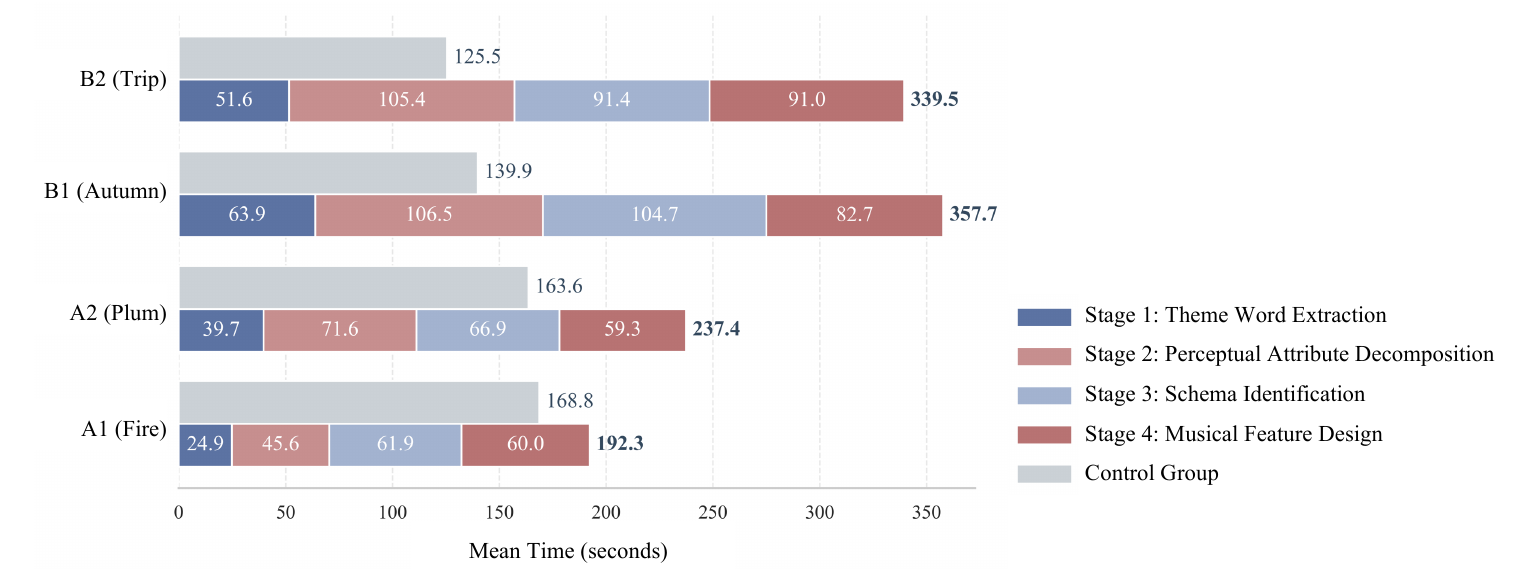}
  \caption{Average ideation time in Study 1}
  \Description{A horizontal stacked bar chart comparing the mean task completion time (in seconds) between the Control Group and Experimental Group across four design themes: A1 (Fire), A2 (Plum), B1 (Autumn), and B2 (Trip). For each theme, the light grey upper bar represents the Control Group's total time, while the lower stacked bar breaks down the Experimental Group's total duration across four distinct ideation stages: Stage 1 (Theme Word Extraction), Stage 2 (Perceptual Attribute Decomposition), Stage 3 (Schema Identification), and Stage 4 (Musical Feature Design). Overall, complex themes required longer ideation times than simple themes in the experimental group (B1 at 357.7 s and B2 at 339.5 s vs. A1 at 192.3 s and A2 at 237.4 s). Across all themes, the Experimental Group spent more total time than the Control Group (e.g., A1: 192.3 s vs. 168.8 s; B1: 357.7 s vs. 139.9 s), with the majority of the additional time allocated to Stage 2 and Stage 3 attribute decomposition and schema mapping.}
  \label{fig:ideation-time}
\end{figure}

The ideation-time statistics are shown in Fig. \ref{fig:ideation-time}. Introducing schemas-based framework was associated with longer ideation processes, particularly for complex themes. This pattern is consistent with the differences observed in their idea descriptions: with structured guidance, participants appeared more likely to elaborate personalized interpretations and less likely to delegate the overall theme to the AI.

\subsection{The Associative Pathways in Music Creation}

\begin{figure}[htb]
  \centering
  \includegraphics[width=0.84\textwidth]{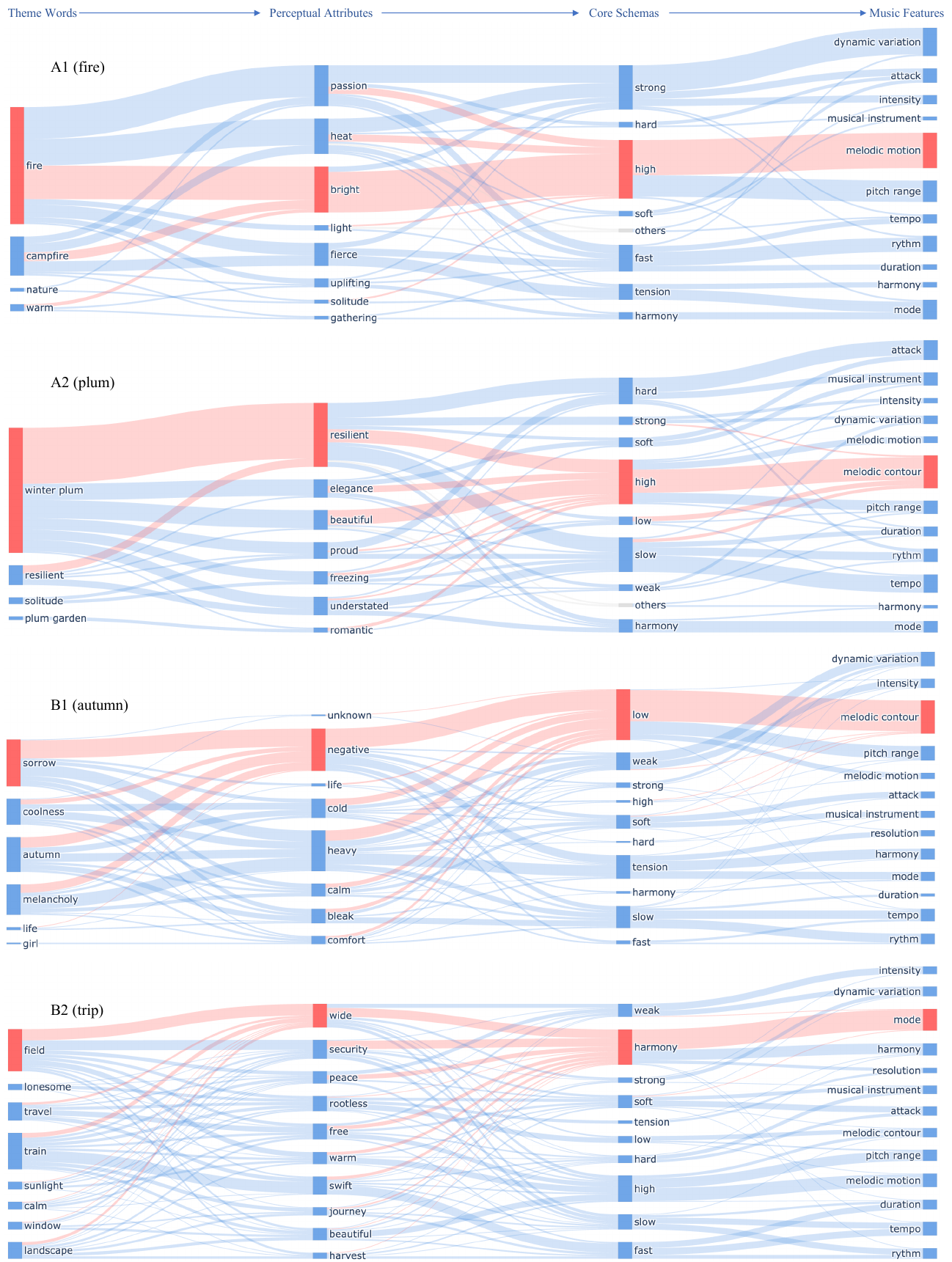}
  \caption{Sankey diagrams of associative pathways across stages for the experimental group in Study 1.}
  \Description{Four-panel Sankey diagram visualizing semantic convergence and divergence across four design stages: Theme Words, Perceptual Attributes, Core Schemas, and Music Features for themes A1 (fire), A2 (plum), B1 (autumn), and B2 (trip). Color-coded ribbons illustrate how concrete theme inputs branch into perceptual descriptors, funnel into bipolar core schemas (e.g., high/low, strong/weak, slow/fast), and terminate in specific acoustic properties.}
  \label{fig:sankey}
\end{figure}

We used Sankey diagrams (Fig. \ref{fig:sankey}) to visualize how users' associative pathways developed. To better reveal the dominant directions of idea flow, semantically similar words were grouped into broader categories; for example, \emph{sad}, \emph{melancholy}, and \emph{somber} were treated as the same semantic category. Across the reasoning process of \emph{theme words → perceptual attributes → core schemas → musical features}, we observed an overall pattern of \emph{convergence → divergence → convergence → divergence}. Convergence was high for theme and schema selection, while perceptual attributes and musical features exhibited greater divergence.

Notably, participants attended more strongly to cultural meanings than to physical properties. For the Theme A2 (plum), most participants focused on the culturally associated quality of \emph{resilience}, rather than on the flower's physical softness or fragility. Similarly, when interpreting the autumn poem, participants primarily attended to the poet's restrained and unspoken emotions, such as \emph{repression}, rather than to the pleasure of autumn weather.

Within the provided schema map, heterogeneous perceptual descriptions consistently converged onto a limited set of schemas, with one to three core schemas accounting for the majority of participants' selections. For example, associations for the Theme A1 (fire) largely converged on \emph{high} and \emph{strong}; Theme A2 (plum) on \emph{high}, \emph{strong}, and \emph{slow}; Theme B1 (autumn) on \emph{low}; and the Theme B2 (trip) on \emph{harmonious} and \emph{high}. This pattern offers one explanation for how music can bridge diverse verbal interpretations: different semantic readings of the same theme may converge onto shared perceptual structures (schemas) that are expressible through music.

However, this convergence was not without exceptions. The term \emph{resilient}, for example, conveys qualities such as persistence and stability in the Chinese cultural context. Different participants associated it with schemas including \emph{hard}, \emph{high}, \emph{slow}, and \emph{strong}. These differences reflect distinct emphases in how participants interpreted the same concept: some focused on an inner sense of calm and stability (\emph{slow}), whereas others emphasized the ability to withstand harsh conditions (\emph{strong}). This phenomenon illustrates a form of perceptual divergence, in which the same description evokes different schemas and different musical features. It may also cause difficulties for a listener to infer the creator's intended meaning from the resulting music alone.

The musical-feature design stage concretized the selected schemas. The diversity of choices at this stage suggests that schemas serve primarily as bridges for association rather than as constraints on individualized creative decisions. Melody-related controls, particularly pitch, were selected most frequently, followed by harmony, timbre, and changes in dynamics. This pattern suggests that participants may perceive these dimensions as especially salient or accessible for shaping musical expression.

\subsection{The Framework Enhances Perceived Control}

We compared the experimental and control groups in Study 1 across three dimensions measured by the subjective evaluation questionnaire: \textbf{Perceived Effectiveness}, \textbf{Perceived Ease of Use}, and \textbf{Perceived Control Enhancement}. The results of the independent-samples \emph{t}-tests are presented in Table \ref{tab:t-tests}. The experimental group scored significantly higher than the control group on both Perceived Effectiveness and Perceived Control Enhancement, suggesting that the schema-mediated framework helped users clarify their creative intents and strengthened their perceived control over the ideation process. No significant difference was found between the two groups in Perceived Ease of Use, but the control group showed a slightly higher mean score. This suggests that the staged guidance did not necessarily make the process easier or more lightweight. Conversely, it tends to achieve a deeper mode of thinking and reflection at the cost of some efficiency and simplicity.

\begin{table}[htb]
\caption{Independent-samples $t$-test results for subjective experience ratings in Study 1 (experimental vs. control group).}
\label{tab:t-tests}
\begin{tabularx}{0.9\linewidth}{@{}Xcccccc@{}}
\toprule
Measure & $t$ & $df$ & $p$ & Mean diff. & SE diff. & Cohen's $d$ \\
\midrule
Perceived Effectiveness & 3.422 & 186 & $<.001$ & 0.303 & 0.089 & 0.499 \\
Perceived Ease of Use & -1.702 & 186 & .090 & -0.250 & 0.147 & -0.248 \\
Perceived Control Enhancement & 2.743 & 186 & .007 & 0.287 & 0.105 & 0.400 \\
\bottomrule
\end{tabularx}
\end{table}

\subsection{Interpretation Improves Understanding of Generated Music}

\begin{figure}[htb]
  \centering
  \includegraphics[width=\textwidth]{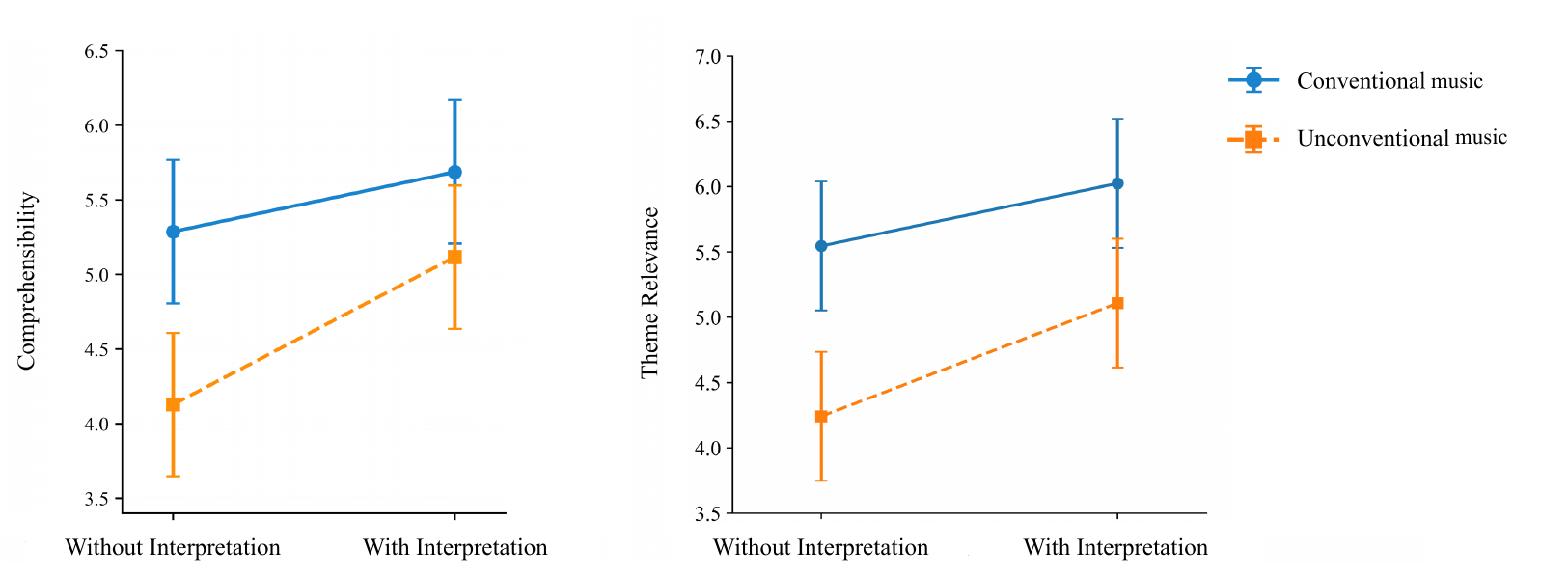}
  \caption{Estimated marginal means (EMMs) of Comprehensibility and Theme Relevance.}
  \Description{An interaction line plot showing the effect of interpretation on music ratings. Both "Conventional music" (solid blue line, increasing from ~5.5 to ~6.0) and "Unconventional music" (dashed orange line, increasing from ~4.2 to ~5.1) receive higher scores in the "With Interpretation" condition compared to "Without Interpretation". Conventional music maintains a consistently higher rating than unconventional music across both conditions.}
  \label{fig:emm}
\end{figure}

The linear mixed-effects model revealed an overall positive effect of \textbf{Interpretation} on participants' evaluations. As shown in Table~\ref{tab:lmm} and Fig.~\ref{fig:emm}, the main effect of Interpretation was significant for both \textbf{Theme Relevance} and \textbf{Comprehensibility}, indicating that providing explanatory information significantly improved participants' understanding of the generated music.

\begin{table}[htb]
\caption{Linear mixed-model analysis results for Theme Relevance and Comprehensibility.}
\label{tab:lmm}
\begin{tabularx}{0.9\linewidth}{@{}lXccc@{}}
\toprule
Dependent variable & Effect & $F$ & $df$ & $p$ \\
\midrule
Theme Relevance & Interpretation & 16.387 & 1, 58 & $< .001^{*}$ \\
& Music Conventionality & 12.426 & 1, 6 & $.012^{*}$ \\
& Interpretation $\times$ Music Conventionality & 3.179 & 1, 412 & $.075$ \\
\addlinespace
Comprehensibility & Interpretation & 18.481 & 1, 58 & $< .001^{*}$ \\
& Music Conventionality & 7.946 & 1, 6 & $.030^{*}$ \\
& Interpretation $\times$ Music Conventionality & 8.324 & 1, 412 & $.004^{*}$ \\
\bottomrule
\end{tabularx}
\end{table}

We further examined whether this effect was moderated by \textbf{Music Conventionality}. A significant \textbf{Interpretation × Music Conventionality} interaction was found for Comprehensibility, \emph{F}(1, 412) = 8.324, \emph{p} = .004, indicating that the positive effect of Interpretation on comprehensibility was more pronounced for unconventional music. For Theme Relevance, the interaction did not reach statistical significance, although the estimated marginal means showed a trend in the same direction. 

\begin{figure}[htb]
  \centering
  \includegraphics[width=0.9\textwidth]{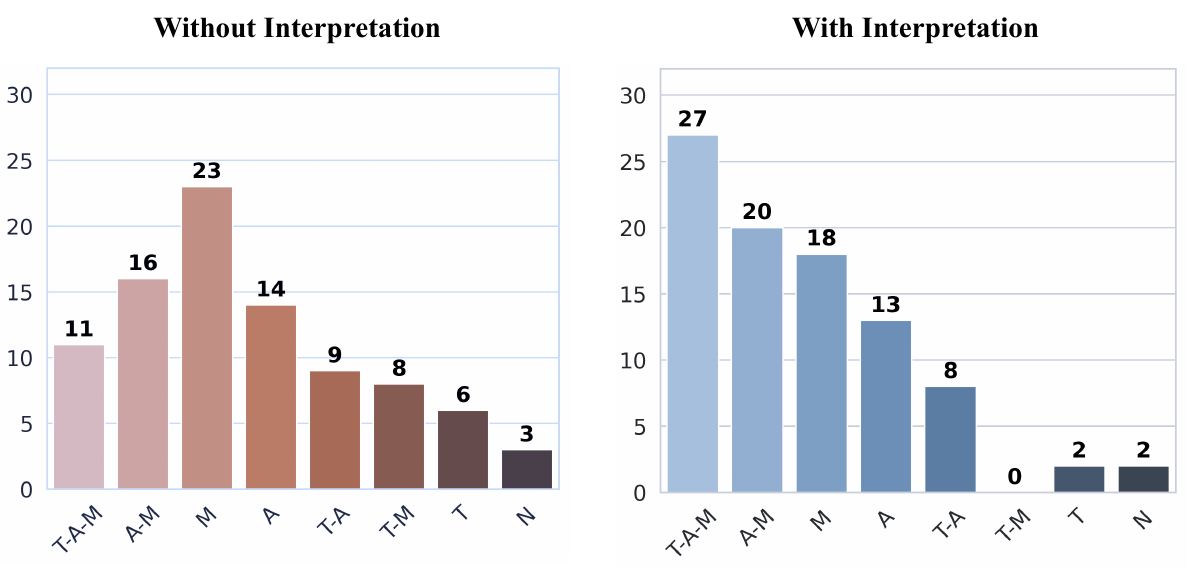}
  \caption{Distribution of revision-suggestion types across the two conditions in Study 2.}
  \Description{Two vertical bar charts comparing the frequency of user revision types across two experimental conditions: "Without Interpretation" (left, reddish-brown bars) and "With Interpretation" (right, blue bars). The x-axis categorizes revisions by modified components: T-A-M, A-M, M, A, T-A, T-M, T, and N (representing combinations of Theme, Attribute, and Music features, or None). In the Without Interpretation condition, single-component adjustments dominate, peaking at M alone (23 occurrences), followed by A-M (16), A (14), and T-A-M (11). In contrast, in the With Interpretation condition, multi-level comprehensive revisions are most frequent, led by T-A-M (27 occurrences), followed by A-M (20), M (18), and A (13). Revisions involving T-M, T alone, and N remain minimal across both conditions (ranging from 0 to 8).}
  \label{fig:revision-types}
\end{figure}

We collected 90 music revision suggestions from each of the interpretation and no-interpretation groups. We coded all suggestions according to the dimensions they addressed: \textbf{T} for theme, \textbf{A} for perceptual attributes, \textbf{M} for musical features, and \textbf{N--} for no revision suggestion. The distribution of coded suggestion types is shown in Fig. \ref{fig:revision-types}.

Participants who received interpretation were more likely to formulate their feedback with a complete \textbf{T--A--M reasoning chain}. A common pattern was to suggest modifying a specific musical feature or element in order to better express a particular attribute of the intended theme. At the same time, these participants were less likely to rely on a direct conceptual leap from the theme (\textbf{T}) to a musical feature (\textbf{M}), such as ``this musical element does not fit the theme.'' This difference suggests that interpretation helps localize what needed to be revised. One characteristic consequence of a \textbf{remote association} is that participants could perceive that something was ``not quite right,'' but had greater difficulty identifying it. The value of structured interpretation lies in decomposing vague holistic judgments into inspectable associations that support more targeted revision.

\section{Discussion}

\subsection{Mediation as Visible Cognitive Scaffolding}
Existing interactive mechanisms in generative AI emphasize goal-oriented processes based on instructions, conditional signals, and step-by-step reasoning. However, artistic meaning often relies on cross-modal, non-linear associations rooted in subjective perception and experience \cite{ref56}. The problem with current models is that they may have already internalized a certain artistic logic internally, but they merely output a determinate ``answer'' without any explanation. Study 2 challenges this paradigm: in the absence of interpretation, users find it more difficult to understand the outcomes, and this lack of understanding leads to unclear revision strategies.

Our schema-based mediation framework does not rely on a set of internalized, opaque mapping logic \cite{ref48}; rather, it serves as a cognitive scaffold for creative associative chains. In contrast to parameterized semantic slider controls \cite{ref2}, we don't force users to quantify their perceptions; instead, we focus on whether the trajectory of creation remains explicitly visible. For users, it supports progressive ideation and creative refinement; for the model, it provides structured data on human internal perception and associative pathways. Unlike predictive paradigms that rely solely on input-output mappings, this visible structure offers a more efficient channel for human-AI empathy.

\subsection{Locating and Diagnosing Associative Divergence}
Mismatches in human-AI collaborative music creation don't always stem from model execution failures, but may also arise from individual differences in users' associative processes. Previous studies have shown that cross-modal associations in music are not fixed one-to-one mappings, but are jointly shaped by differences in emotion, experience, familiarity, and other factors \cite{ref27,ref32,ref28}. However, in traditional prompt-based interaction modes, users can only observe the inputs and generated outputs, without being able to examine such associative discrepancies.

Theme A2 (plum) illustrates this phenomenon. Most participants consistently regarded ``resilience'' as the most important attribute, yet interpreted it through different schemas, such as \emph{strong} and \emph{slow}. These interpretations are not simply right or wrong, but rather reflect different perceptual tendencies toward the same concept, which lead to divergent musical feature choices. Therefore, our framework is not intended to eliminate such cognitive divergence, but to make it localizable and negotiable. Compared with presupposing a single correct mapping, externalizing uncertainty better aligns with the authentic cognitive process of music creation.

Participants in Study 2 who received explanatory information rated the generated music more highly and were better able to propose revision suggestions based on the complete reasoning chain. \textbf{Therefore, we define controllability in artistic generation as diagnosability}, i.e., whether users can understand why and at which stage a creative divergence occurs. Human--AI co-creative systems should attend to the cognitive origins of users' control actions, rather than merely focusing on the terminal end of signal input.

\subsection{From AI Alignment to Self-Intent Deconstruction}
In mainstream discussions of intent alignment, there exists an implicit assumption that users hold a predefined goal, and the system's task is merely to match it as accurately as possible \cite{ref64,ref65,ref66}. However, in open-ended practices such as music creation, intent rarely exists in a fully formed state beforehand. It often emerges and takes shape gradually through the iterative process of expressing, listening, comparing, and revising \cite{ref52,ref42}.

We attempt to provide a cognitive support pathway in this regard, not by reading a complete intent, but by decomposing the narrative into associative chains, guiding users to make judgments and choices at each step. This process is essentially a deconstruction of user's own creative intent: they need to continually examine ``what exactly do I want to express'' and ``which musical features correspond to this perceptual attribute.'' In this process, intent is no longer a priori input, but is progressively externalized into actionable creative instructions through users' ongoing reflection and adjustment on the intermediate associations.

In other words, creative alignment should consist of two progressive stages: first, leveraging explicit associative pathways to help users clarify their own creative intents, and then having the AI understand and execute these already clarified goals. This perspective also implies that a co-creative system needn't anchor itself to some ``correct'' understanding of intent. It should preserve sufficient space for reflection for different users, supporting them in autonomously constructing and maintaining diverse associative pathways.

\subsection{Multi-Stage Ideation Supports Personalized Creative Expression}
AI can rapidly generate creative outputs, but this efficiency may also come with a hidden cost: compressing users' personalized exploration and reflection. Previous studies have shown that although generative AI can improve the creativity scores of individual works, it may reduce the overall diversity of outputs \cite{ref16}; its mode of participation can also significantly affect users' critical engagement and autonomy \cite{ref62,ref63}.

The results of Study 1 are consistent with this. When faced with complex themes, the descriptions of participants in the control group mostly remained at a broad level that was semantically close to the theme but lacked deconstruction and transformation. Whereas participants in the experimental group took longer in their ideation, they also produced final descriptions with more specific musical features and richer cultural and emotional interpretations. Ideation time alone is not a sufficient indicator of deep reflection, but when considered in conjunction with the content of the descriptions, the difference in the experimental group lies in their more thorough engagement with the associative reasoning process of ``theme → perceptual attributes → musical parameters,'' rather than jumping directly to the outcome.

This suggests that our framework provides effective intermediate anchors, helping users re-engage with the creative translation process that might otherwise be entirely delegated to the AI. This engagement sacrifices, to some extent, the immediacy of operation, yet facilitates deeper thinking. For example, when faced with the same complex theme, the experimental group actually produced more diverse creative interpretations. This suggests that AI co-creative systems should not take convenience as their sole design principle, but should also support users in engaging in more deliberative, personalized, and reflective creative thinking.

\subsection{Limitations and Future Work}

The framework was evaluated through a \textbf{conceptual simulation} rather than a fully implemented interactive system. Participants followed the proposed framework to develop their ideas, while the mapping from schemas to musical features was not connected to the latent space of a generative model. The current findings therefore primarily reflect the \textbf{cognitive guidance effects of the framework}, rather than the usability of any particular system interface. Future work will implement the framework as a functional interactive system and evaluate whether its effects transfer to real-world co-creative settings.

Second, the effectiveness of the interaction framework was evaluated within a specific cultural context, and its cross-cultural applicability remains to be established. The current study may not capture differences in associative habits across cultural backgrounds. Future studies should therefore include more culturally diverse samples to examine the transferability and boundary conditions of the framework. In addition, this work primarily focuses on music medium. Whether schema mediation can provide similar benefits in other abstract creative domains remains an open question.

Schema representation and staged procedure are intrinsically coupled: decomposing an association chain into layers is itself a staged way of thinking, so we don't attempt to separate their contributions. What remains open are intervention-level questions, namely instructional richness, ideation time, and vocabulary priming, which future studies with matched control materials and time measures can address. Future research may also examine the transferability of schema-mediated mechanisms, their long-term effects, and culturally situated association maps. 

\section{Conclusion}

Individual differences in association mean there is no single correct mapping from story to musical parameters. The primary obstacle to aligning intent in human-AI co-creation lies in the invisibility of associative divergences. This study introduces schemas as a mediator between narrative and music, constructing a cognitive framework that transforms invisible remote associations into inspectable, staged decisions. Two experiments validated two predictions: in the forward pathway, the framework guided users to produce clearer, more executable creative descriptions; in the reverse pathway, the framework's interpretive strategy significantly improved users' comprehension and evaluation of the generated music. At the same time, structured guidance did not constrain creative freedom, but enhanced users' perceived controllability. 

Despite limitations in technical implementation and sample scope, this study offers a new design strategy for human--AI co-creative systems: externalizing associative pathways to make them inspectable and negotiable. In essence, this allows both users and AI models to deconstruct their own intentions within a shared framework, making their thinking visible to each other, and thereby enabling a higher level of synchronization. We envision AI's role not as a replacement for human creators, but as a catalyst for deeper creativity and expression.

\bibliographystyle{ACM-Reference-Format}
\bibliography{sample-base}

\end{document}